\documentclass[%
 reprint,
 amsmath,assume,
 aps,
]{revtex4-2}
\usepackage{datetime2}
\usepackage{xcolor}
\usepackage{amsmath}
\usepackage{subfigure}
\usepackage{graphicx}
\usepackage{dcolumn}
\usepackage{bm}
\usepackage{hyperref}
\usepackage{epstopdf}

\usepackage[export]{adjustbox}
\begin{document}

\preprint{APS/123-QED}

\title{Anisotropic magneto-photothermal voltage in ${Sb_2}{Te_3}$ topological insulator thin films.}
\thanks{A footnote to the article title}%

\author{Subhadip Manna}
 \altaffiliation{Department of Physical Sciences, IISER Kolkata}
 \author{Sambhu G Nath}
 \altaffiliation{Department of Physical Sciences, IISER Kolkata}
 \author{Samrat Roy}
\altaffiliation{Department of Physical Sciences, IISER Kolkata}
 \author{Soumik Aon}
 \altaffiliation{Department of Physical Sciences, IISER Kolkata}
 \author{Sayani Pal}
 \altaffiliation{Department of Physical Sciences, IISER Kolkata}
  \author{Kanav Sharma}
  \altaffiliation{Department of Physical Sciences, IISER Kolkata}
 \author{Dhananjaya Mahapatra}
  \altaffiliation{Department of Physical Sciences, IISER Kolkata}

 \author{Partha Mitra}
 \altaffiliation{Department of Physical Sciences, IISER Kolkata}
 \author{Sourin Das}
 \altaffiliation{Department of Physical Sciences, IISER Kolkata}
 \author{Bipul Pal}
 \altaffiliation{Department of Physical Sciences, IISER Kolkata}
\author{Chiranjib Mitra}%
 \email{Corresponding author:chiranjib@iiserkol.ac.in}
\affiliation{%
 Indian Institute of Science Education and Research Kolkata,
Mohanpur 741246, West Bengal, India
}%

\date{\today}

\begin{abstract}
We studied longitudinal and Hall photothermal voltages under a planar magnetic field scan in epitaxial thin films of the Topological Insulator (TI) $Sb_2Te_3$, grown using pulsed laser deposition (PLD). Unlike prior research that utilized polarized light-induced photocurrent to investigate the TI, our study introduces advancements based on unpolarized light-induced local heating. This method yields a thermoelectric response exhibiting a direct signature of strong spin-orbit coupling. Our analysis reveals three distinct contributions when fitting the photothermal voltage data to the angular dependence of the planar magnetic field. The interaction between the applied magnetic field and the thermal gradient on the bulk band orbitals enables the differentiation between the ordinary Nernst effect from the out-of-plane thermal gradient and an extraordinary magneto-thermal contribution from the planar thermal gradient. The fitting of our data to theoretical models indicates that these effects primarily arise from the bulk states of the TI rather than the surface states. These findings highlight PLD-grown epitaxial topological insulator thin films as promising candidates for optoelectronic devices, including sensors and actuators. Such devices offer controllable responses through position-dependent, non-invasive local heating via focused incident light and variations in the applied magnetic field direction. 

\end{abstract}

\maketitle


\section{\label{sec:level1}Introduction }

Topological insulators(TI) are emerging quantum materials that have garnered significant attention over the past decade and a half due to the nontrivial topology of electronic wavefunctions in the bulk of the material\cite{moore2010birth}\cite{zhang2009topological}\cite{hasan2010colloquium}. Back scatterings suppressed metallic surface states in TI, which have Dirac-like dispersion protected by the time-reversal symmetry, allow a directional flow of surface spins that are locked with their momentum \cite{ando2013topological}\cite{hasan2011three}. These unique properties of the TI add new dimensions to the field of spintronics and optoelectronics\cite{he2019topological}\cite{politano2017optoelectronic}. Anisotropic magnetoresistance (AMR) and planner hall effect (PHE) are characteristic signatures of ferromagnetic materials that have been widely acknowledged for a long time\cite{1058782}\cite{ky1967planar}\cite{smit1951magnetoresistance}\cite{campbell1970spontaneous}. Recently the same phenomenon has been observed in topological insulators\cite{taskin2017planar}\cite{budhani2021planar} along with Weyl and Dirac semimetals\cite{kumar2018planar}\cite{li2018giant}, which are non-magnetic in nature. Until now, it has been widely acknowledged that the AMR and the PHE in topological semimetals result from the chiral anomaly effect inherent in these systems \cite{nandy2017chiral}. However, in topological insulators (TI), this matter remains unsettled. Taskin $et$ $al.$ \cite{taskin2017planar} have given a minimal model to capture their experimental observation, where they have addressed that the magnetic field-induced anisotropic lifting of surface protection encourages spin-flip scattering from random impurities which gives rise to AMR and PHE in $Bi_{2-x}Sb_xTe_3$ TI. Whereas Suleav $et$ $al.$ \cite{sulaev2015electrically} have considered coupling of both top and bottom surface states through side surfaces and the bulk to justify the AMR observed in $BiSbTeSe_2$ nanoflake. According to their physical model, an in-plane magnetic field induces a shifting of the two Dirac cones on the top and the bottom surfaces, which are coupled through the side surfaces and bulk, generating a net spin accumulation along the field that produces the anisotropy. However, Nandy $et$ $al.$ \cite{nandy2018berry} considered the effect due to nontrivial Berry curvature and the orbital magnetic moments in the bulk conduction limit to explain negative longitudinal magnetoresistance (LMR) and PHE in TI. Also, the hexagonal warping effect is considered to explain the in-plane AMR as well as out-of-plane variation as mentioned by Akzyanov $et$ $al.$ \cite{akzyanov2018surface} where a three-fold variation of PHE has been observed. Zheng $et$ $al.$ \cite{zheng2020origin}considered in-plane magnetic field-induced band tilting which deforms the fermi surface and thereby allows the backscattering in topological surface states as an origin of PHE and AMR in TI.\\

So far studies have focused more on magnetoelectric transport and less attention has been paid to the magnetothermal properties, despite the fact that TI materials like ${Bi_2}{Se_3},$${Bi_2}{Te_3},$${Sb_2}{Te_3}$, etc. are known as excellent thermoelectric candidates\cite{imamuddin1972thermoelectric}\cite{xu2017topological}\cite{ivanov2018thermoelectric}\cite{osterhage2014thermoelectric}\cite{das2015defect}. Recently a group has predicted asymmetric magnon scattering induced unidirectional Seebeck effect in magnetic TI (MTI)/TI heterostructure \cite{yu2019unidirectional} and also hexagonal warping mediated nonlinear planar Nernst effect in nonmagnetic TI \cite{yu2021hexagonal}. In another recent experiment thermal spin current has been observed in ${Bi_2}{Se_3}/{CoFeB}$ heterostructure, understood as the spin Nernst effect (SNE), which is nothing but the thermal analog of the spin Hall effect(SHE) \cite{jain2022thermally}.\\

In this work, we address the magnetic field-dependent anisotropic photothermal voltage in $Sb_2Te_3$ thin films measured at room temperature. Unlike prior studies that rely on polarized light for studying the TI, we introduce a new method based on laser-induced local heating. In this novel approach, we developed a technique wherein unpolarized light efficiently induces local heating. This, in turn, leads to a thermoelectric response showing three distinct contributions when fitting to the angular dependence of the magnetic field. Our method has the advantage of controllability of moving the laser spot along the length of the sample in a non-invasive manner, which cannot be attained by Ohmic heat contacts. Our measurements are based on Lock-in detection, where we were able to measure thermally generated voltage. The angular dependence of photothermal voltage on the planar magnetic field and the resulting analysis that separates the three different components throw new light into the topological nature of the bulk bands, probed through temperature fields. 

\section{\label{sec:level2}Methods}

${Sb_2}{Te_3}$ films of a thickness of 50nm were grown on ${SiO_2}$ (300nm)/Si substrate using pulsed laser deposition (PLD) technique. The thickness and surface morphology of the films were characterized by atomic force microscopy (AFM)(Supplementary Fig.1(d)). The cross-bar structure of the device was patterned by a standard electron-beam lithography process using PMMA as the electron beam resist. The target material used for film growth consists of {99.999\%} pure ${Sb_2}{Te_3}$ pellet. The films were grown through ablation of the target using KrF excimer laser (wavelength 248nm) with a laser pulse frequency of 1Hz. A base pressure of $6\times10^{-6}$ mbar was attained before the deposition and a partial pressure of $5\times10^{-1}$ mbar of flowing Argon was maintained throughout the deposition. The optimized substrate temperature of $230^\circ$C and laser fluence around 0.9 J$cm^{-1}$ ensured high-quality film growth, confirmed by X-ray diffraction (XRD)(Supplementary Fig.1(a)) and Field emission scanning electron microscopy (FESEM)(Supplementary Fig.1(c)). After the deposition, films were annealed for 30 minutes at the same temperature to achieve high crystallinity and reduced surface roughness. Contact electrodes were patterned using a standard photolithography process followed by electron-beam evaporation of Cr(6nm)/Au(30nm).

\begin{figure*}
    \centering
    
        \includegraphics[scale=0.6]{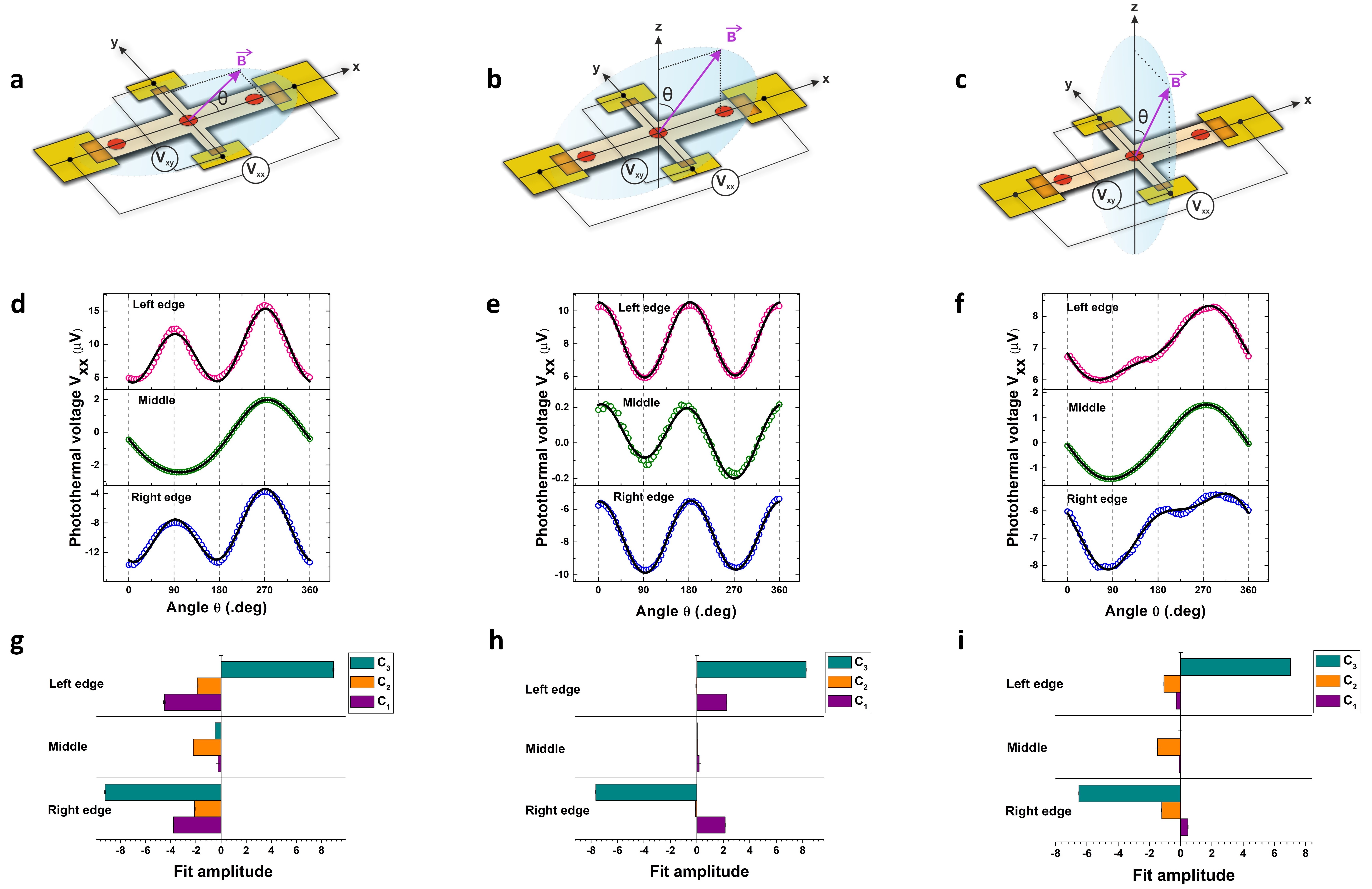}

\caption{\label{fig:wide}\textbf{photothermal voltage in three different scanning geometries :} Geometries of longitudinal and transverse photothermal voltage measurements with rotating magnetic field (0.3T) \textbf{B} in the x-y [\textbf{a}],x-z [\textbf{b}] and y-z [\textbf{c}] planes. For in-plane (x-y) scanning, the angle $\theta$ is measured from the x-axis, and for out-of-plane scanning, $\theta$ is measured from the z-axis. Red circles denote the laser spot position (left edge, middle, right edge) as illustrated in schematics of [\textbf{a}],[\textbf{b}], and [\textbf{c}]. [\textbf{d,e,f}] measured photothermal voltage as a function of magnetic field angle for the scanning geometries x-y,x-z, and y-z planes. Red, green, and blue curves represent measured photothermal voltage for laser spot position left edge, middle, and right edge respectively. The black solid line shows the decomposition of two angle-dependent signals of $sin{\theta}$ and $cos{2\theta}$. [\textbf{g,h,i}] Comparison of fitting parameters of Eq.\eqref{e:1} for x-y, x-z and y-z geometries respectively. }

\end{figure*}

\section{\label{sec:level3}Experimental observations} 
        Photothermal voltage has been measured in a cross-bar device by scanning the magnetic field (0.3T) in three different geometrical planes x-y, x-z, and y-z as shown in Fig.1(a-c). For in-plane (x-y) scanning, the angle $\theta$ is measured from the x-axis, and for out-of-plane scanning, $\theta$ is measured from the z-axis as shown in Fig.1(a-c).  In each measurement geometry, three different spot positions have been chosen where the thermal gradient is generated by a focused laser (30mW) of 150$\mu$m diameter. One spot is positioned at the middle of the sample and the other two spots are close to the left and right edges. Hence, laser illumination at these positions generates distinct thermoelectric potentials across the sample. Fig.1(d-f)  refers to the measured photothermal voltage as a function of the magnetic field angle for the aforementioned measurement configurations. In the x-y scanning of the magnetic field, when the laser is illuminated in the middle of the sample, measured photothermal voltage shows a single peak/dip with the variation of the magnetic field angle. In contrast, near the edges, two peaks/dips variation is observed with asymmetric amplitudes. On the other hand, in the x-z scan, two peaks/dips can be seen as a function of the magnetic field angle, independent of the laser spot positions. It is seen that the amplitudes corresponding to the two peaks/dips for both the left and right edges are quite symmetric. However, with the laser spot in the middle, an asymmetric nature is observed, but the magnitude is very small in comparison to the edges. Despite this, a single peaks/dips nature is observed in the y-z scan, except for a slight asymmetry when the laser spots are near the edge.\\

        The above observations of photothermal voltage variation with magnetic field angle arise from the thermal gradient generated by the laser illumination. In the middle of the sample, net in-plane thermal gradient vanishes and only an out-of-plane thermal gradient along the thickness is worthy of consideration. Moving close to the left and right edges both in-plane and out-of-plane thermal gradients contribute to the generation of photothermal voltage. Considering these, the longitudinal photothermal voltage equation is formulated as:\\
    \begin{eqnarray} \label{e:1} 
    V_{xx} = C_1 cos{2\theta} + C_2 sin{\theta} + C_3
    \end{eqnarray}
    where the first term in the above equation represents anisotropic photothermal voltage due to in-plane thermal gradient (${\nabla}T_{x}$) acting along the x-direction, and the second term comes from the perpendicular temperature gradient (${\nabla}T_{z}$) along the z-direction. $C_3$ is the parameter independent of the field rotation angle. Black solid lines in Fig.1(d-f) are fitted curves using  Eq.\eqref{e:1}, which matches the experimental findings. For each scanning geometry, the amplitude of fitted components is presented in the bar diagram in Fig.1(g-i) for the left, right, and middle illumination positions. It is evident that, for the x-y scan, both in-plane and out-of-plane thermal gradients contributed significantly near the edges. Whereas for the x-z scan, only the in-plane thermal gradient and for the y-z scan only the out-of-plane thermal gradient play an important role in the observed photo thermal voltage. \\
    
  Microscopic understanding is needed to explain the experimental observations. The second term in Eq.\eqref{e:1} has an origin from the ordinary Nernst effect (${\nabla}T_{z}\times B$)\cite{aono1970nernst}\cite{barnard1974thermoelectricity}. The $sin{\theta}$ signal is a result of the perpendicular temperature gradient ${\nabla}T_{z}$ produced by laser irradiation along the sample thickness. The ordinary Nernst effect generates a voltage along the transverse direction (x-axis) for the x-y and y-z scan  Fig.1(a, c, g, i) and along the y-axis for the x-z scan Fig.1(b, h). In general, laser irradiation generates an observable photocurrent (photovoltage) in TI which includes both surface and bulk excitations. However, circularly polarized light can selectively excite the surface state and generate a spin-polarised helical photocurrent\cite{mciver2012control}\cite{pan2017helicity}. In earlier work, we have shown that helicity can be tuned by changing the polarization from left circular to right circular as well as changing the incident angle of light \cite{roy2022photothermal}. By applying a planar magnetic field these excitations will get modified and the phenomena behind this is known as the magneto-gyrotropic photogalvanic effect (MPGE) \cite{weber2008magneto}\cite{junck2013photocurrent}, a sinusoidal variation of photovoltage is observed with magnetic field angle while keeping light polarization fixed\cite{chen2021plane}.In our case ordinary Nernst effect is the key mechanism of sinusoidal variation of photovoltage as confirmed by thickness dependency (see supplementary fig)  But MPGE is insufficient to explain the cos2$\theta$  modulation (first term in Eq.\eqref{e:1})of the observed photothermal voltage.\\
        
   Let us consider how the magnetic field affects the transport of photogenerated carriers under the influence of the thermoelectric potential. Despite the fact that TI has a distinct bulk band structure, surface states in TI have received more attention over the bulk due to their unique spin-momentum locking and topological protection against backscattering. However, in our case, the photo-excited carriers in ${Sb_2}{Te_3}$ primarily originate from bulk bands, and the effect of bulk states cannot be avoided. Additionally, we are not using any circularly polarized light, and that reduces the spin-selective surface sate excitation and its contribution. Furthermore, the fact that the intrinsic fermi level is located inside the bulk valance band encourages us to focus on bulk states rather than surface states (Supplementary Note 8).
   Recent observation confirms the existence of negative longitudinal magnetoresistance (LMR) and planar hall effect (PHE) in TI. The chiral anomaly effect results in negative LMR and PHE in topological semimetals like Weyl and Dirac semimetals, but there is no such well-defined chiral anomaly effect in TI. Recent theoretical and experimental works have demonstrated that nontrivial Berry phase effects of bulk bands can result in negative LMR in topological insulators even in the absence of chiral anomaly \cite{nandy2018berry}. By solving the Boltzmann transport equation semi-classically, an \textbf{E.B} term resembling a topological term emerges (Supplementary Note 1) \cite{nandy2018berry}\cite{morimoto2016semiclassical}. This can be understood as a semi-classical representation of chiral anomaly in TI in the presence of non-trivial Berry flux through the fermi surface whereas it has purely quantum origin in topological semimetals. The simplest representation of photothermal voltage (photocurrent) can be deduced as \\
    \begin{eqnarray} \label{e:2} 
    \textbf{V}=\beta (\nabla\textbf{T}\cdot \textbf{B})\textbf{B}
    \end{eqnarray} 
 where $\nabla\textbf{T}$ and \textbf{B} are thermal gradient induced by laser illumination and externally applied magnetic field respectively. Throughout the experiment, no external electric bias has been applied. However, a thermoelectric potential gradient (field) is created by laser illumination and it has an equivalent form of an external field $\textbf{E}= -S\cdot\nabla\textbf{T}$\cite{barnard1974thermoelectricity}\cite{san1970nernst}, where $S$ is the Seebeck coefficient of the material.
    
\begin{figure}
    \centering
    \includegraphics[width=1\linewidth]{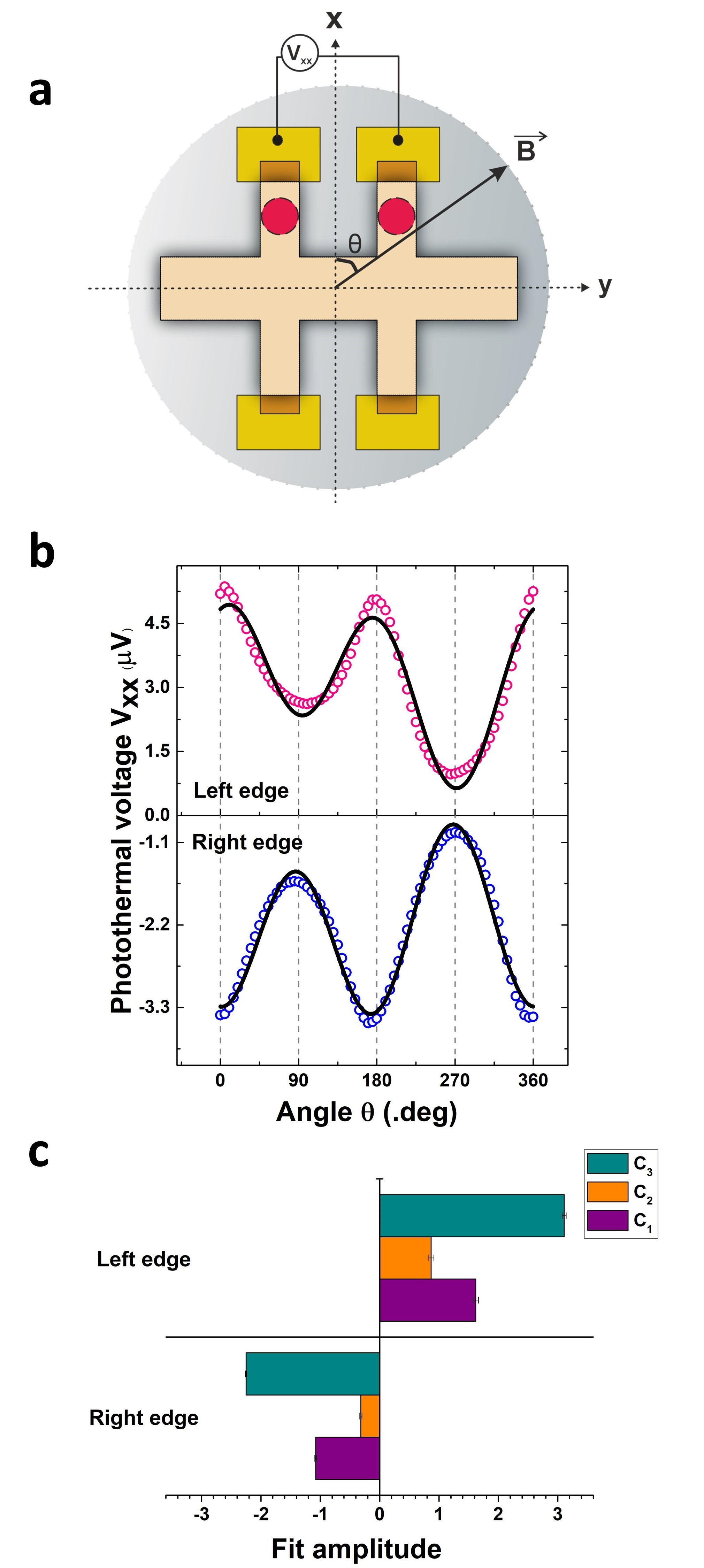}
    
    \caption{\label{fig:wide}\textbf{photothermal voltage measurement in hall bar device :} [\textbf{a}] longitudinal photothermal voltage measurement with rotating magnetic field (0.3T) \textbf{B} in the x-y plane. Red circles denote the laser spot
position (left edge, right edge) as illustrated in the schematics.[\textbf{b}]Observed photothermal voltage as a function of magnetic field angle for the scanning geometry x-y. The red and blue curves represent measured photothermal voltage for the laser spot position on the left edge and right edge respectively. Black solid
lines are fitted curves using Eq.\eqref{e:1}.[\textbf{c}] Comparison of fitting parameters of Eq.\eqref{e:1} for both laser spot positions .}
    
\end{figure}
    
    In our measurement scheme, the in-plane thermal gradient is always directed along the x-axis and changes sign when the laser illumination spot is moved from one edge to the other (cf.Fig.1(a-c)). In this way, the longitudinal photothermal voltage for the three different scanning planes (x-y,y-z,x-z), by considering positive thermal gradient, can be expressed as \\
    \begin{eqnarray}  \label{e:5}
    V_{xx}(xy)=\beta\nabla{T}_x B^2cos^2(\theta)
    \end{eqnarray}
     \begin{eqnarray}  \label{e:6}
    V_{xx}(xz)=\beta\nabla{T}_x B^2sin^2(\theta)
    \end{eqnarray}
     \begin{eqnarray}  \label{e:7}
    V_{xx}(yz)=0
    \end{eqnarray}

      The above three equations can be truncated as $V_{xx}(xy,xz,yz)=C_1 cos(2\theta)$ (for detailed calculation see Supplementary Note 2), which is the first term in Eq.\eqref{e:1}.\\

      It is clear from the above discussion that changing the in-plane thermal gradient from positive to negative by moving the laser illumination spot from one edge to the other will lead to a change in the sign of parameter $C_1$. However, experimental data presented in Fig.1(g,h) contradicts the theoretical aspects discussed so far, no change of sign of parameter $C_1$ is observed. This discrepancy between theory and experiment can be resolved by incorporating laser-induced local heating. In the elongated sample, ${\nabla}T_{x}$ between the two ends of the sample is not significantly generated by the laser heating which is somewhat local\cite{kim2017observation}. Considering the effect of different edges (left and right) individually, Eq.\eqref{e:2} can be modified as\\
\begin{eqnarray} \label{e:8} 
    \textbf{V}'=(\textbf{V}_{Left}-\textbf{V}_{Right})
    \end{eqnarray}
     For in-plane field rotation (x-y scan) Eq.(8) reduces to \\
     \begin{eqnarray}  \label{e:9}
    \textbf{V}'_{xx}(xy)=\beta B((\nabla\textbf{T} \cdot \textbf{B})_{Left}-(\nabla\textbf{T} \cdot \textbf{B})_{Right})cos(\theta)
    \end{eqnarray}
When the laser is directed towards the left edge, let's assume a positive thermal gradient near this edge, while no thermal gradient is present near the right edge. Similarly, when the laser is focused near the right edge, only a negative thermal gradient occurs in that vicinity. Both these cases yield the same outcome as per Eq.\eqref{e:8}, regardless of the presence of positive or negative thermal gradients. Consequently, the sign of parameter $C_1$ remains unaltered in both scenarios.  In principle, our experimental findings accurately depict this phenomenon.\\
\begin{figure}
    \centering
    \includegraphics[width=1\linewidth]{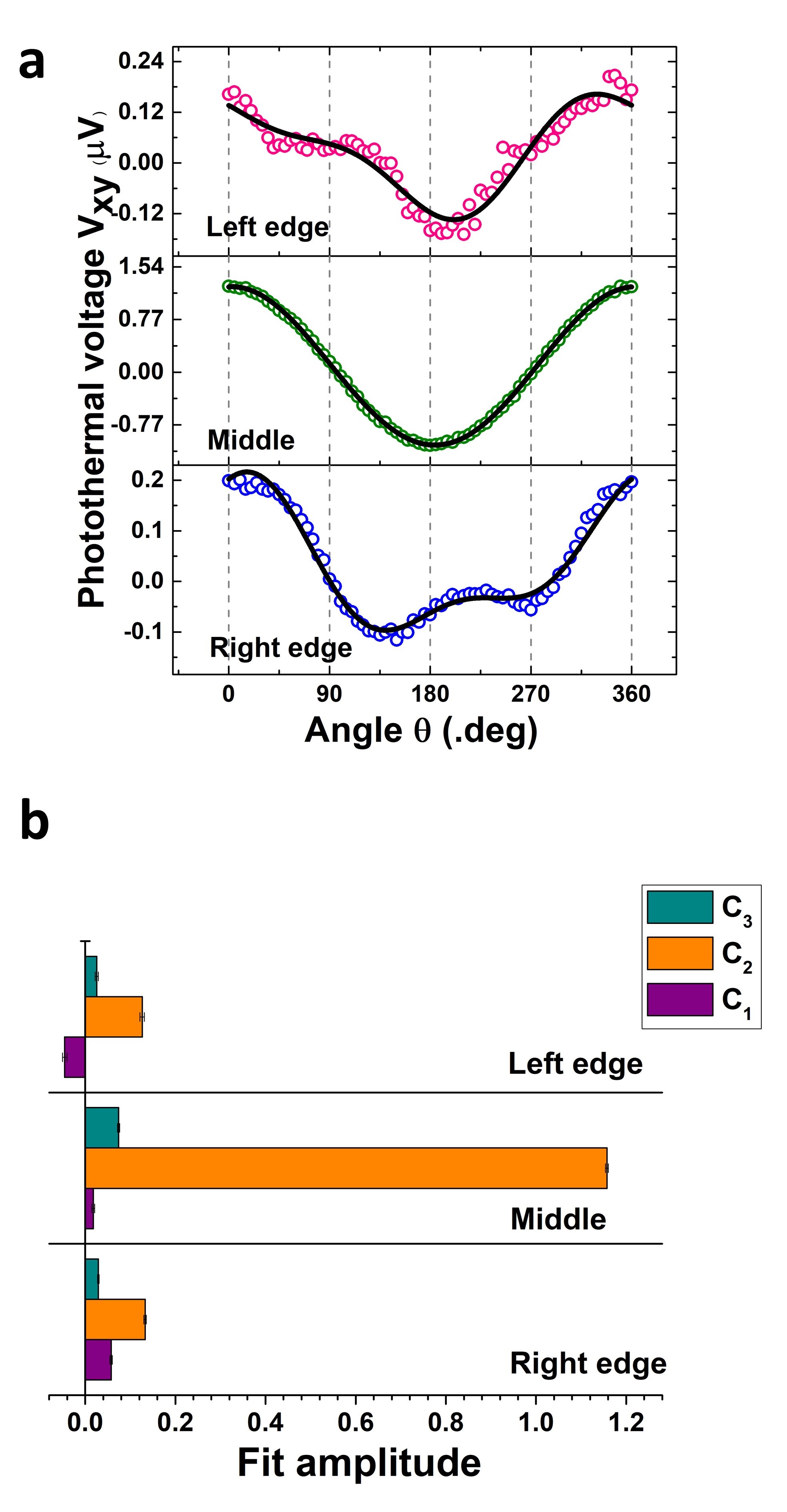}
    \caption{\label{fig:wide}\textbf{Transverse photothermal voltage:} 
 [\textbf{a}]Observed transverse photothermal voltage as a function of magnetic field angle for the scanning geometry x-y (Fig 1(a)). Red, green, and blue curves represent
measured photothermal voltage for laser spot position at the left edge, middle, and right edge respectively. Black solid
lines are fitted curves using Eq.\eqref{e:10}.[\textbf{b}] Comparison of fitting parameters of Eq.\eqref{e:10} for different laser spot positions. }
    
\end{figure}

To confirm the microscopic assumption of localized heating near each contact separately, a Hall bar-shaped device has been fabricated. photothermal voltage measurements were conducted by directing the laser spot toward the top left and top right contacts of the Hall bar sample, as depicted in Fig. 2(a) while rotating the magnetic field within the x-y plane. The experimental observations are depicted in Fig. 2(b), where red and blue open circles illustrate the observed photothermal voltage near the left and right edges, respectively. The black lines represent the fitted curves by employing Eq.\eqref{e:1}. A comparison of the corresponding fitting parameters is presented in the bar diagram shown in Fig. 1(c). It is clear that not only the $C_3$ value (which is independent of field rotation) but also  $C_1$ and $C_2$ parameter flip their sign as one changes the laser spot position from the top left edge to the top right edge. This is expected if laser heating generates locally confined photothermal voltage near each contact which we have assumed. This is because $\nabla\textbf{T}$ is always directed along the positive x-axis for each laser spot position in the Hall bar geometry. As a result $(\nabla\textbf{T} \cdot \textbf{B})$ always has the same sign irrespective of the top left and top right edge laser illumination. That leads to a sign change in $\textbf{V}'_{xx}(xy)$ in Eq.\eqref{e:9} (equivalent to sign change of parameter $C_1$ of Eq.\eqref{e:1}.\\

This report places its primary focus on the anisotropic longitudinal photothermal voltage rather than the transverse Hall response. However, it's worth noting that a planar Hall voltage, which is another intriguing property originating from the non-trivial band topology in TI, can also be observed. To measure the transverse photothermal voltage, the same measurement configuration has been followed as depicted in Fig. 1(a) (x-y field scan). However, the photothermal voltages were measured along the transverse direction (along the y-axis) for three different laser spot positions (left, right, and middle). The observed results are plotted as open circles, as shown in Fig 3(a). The black solid lines indicate the fitting with the equation\\ 
\begin{eqnarray}  \label{e:10}
    V_{xy} = C_1 ^\prime sin{2\theta} + C_2 ^\prime cos{\theta} + C_3 ^\prime
\end{eqnarray}
    The first term originates from the planar Hall effect due to the in-plane thermal gradient ($\nabla{T}_x$), while the second term represents the ordinary Nernst term. The fitting parameters are presented using bar plots as shown in Fig.3(b). From Fig. 3(a,b), it is evident that a substantial photothermal Hall voltage is observed only when the laser is shined at the middle position and predominantly comes from the ordinary Nernst effect. This is expected because when the laser is illuminated in the middle there is no net in-plane thermal gradient and only an out-of-plane thermal gradient is present. On the other hand, when the laser has been focused near the left and right edge planner photothermal gradient comes into play and a reasonable planner hall response has been observed along with Nernst voltage. Also, a sign change of the planner hall coefficient has been observed with the change of thermal gradient. However, planner hall amplitudes are very small compared to longitudinal photothermal voltage as shown in Fig.1(d,g). Which again confirms the laser-induced local heating. As long as the laser is illuminating inside the hall channel(in the middle spot) large photothermal hall voltage has been observed and when illuminated outside the channel(near the left and right edge) voltage has been suppressed significantly.\\

The photo-voltage measurements on epitaxial thin films of $Sb_2Te_3$, grown using PLD, reveal intriguing angular dependence in response to a planar magnetic field scan. This photo-voltage is proposed to emerge from the diffusion of charges under the thermal gradient resulting from the local heating generated by the laser spot. In general, the observed dependence of the longitudinal photo-voltage on the magnetic field angle can be attributed to three distinct components. The first component, $C_1$, emerges from a thermal gradient along the length of the sample, while another component, $C_2$, results from the out-of-plane thermal gradient (along the thickness). There is a third term, $C_3$, that remains unaffected by the angle of rotation of the magnetic field. All contributions were determined by varying the magnetic field's plane of rotation across three distinct planes, x-y, y-z, and z-x, and also by positioning the laser spot at three different positions (left, right, and middle) of the sample. We fitted all nine sets of data (comprising three positions for each of the three different planes) using the photo-voltage Eq.\eqref{e:1}, thus ascertaining the three different components. The magnitude and the sign variation of component $C_1$ on the angular field as a function of the position of the laser spot, for a particular plane of rotation corroborates that the dependence is due to the non-trivial coupling of the in-plane thermal gradient and the magnetic field. The term $C_2$, arising from the thermal gradient across the sample's thickness, was ascertained by fitting the angular field-dependent data obtained from samples with various thicknesses. We have also managed to adjust the laser spot's position with an accuracy of 10$\mu$m, thereby allowing precise control over both the thermal gradient and the magnitude and polarity of the photothermal voltage. This could enable the rapid switching of the photo-voltage in a microscopic device, thereby allowing us to utilize the device as a switch. This work demonstrates that the  \textbf{E.B} term can be activated even by the thermal gradient generated via laser-induced heating, except that here it is the non-trivial spin-orbit coupled bulk bands that come into play, as predicted by theoretical works\cite{nandy2018berry}. Our findings imply that epitaxial thin films of TI materials like $Sb_2Te_3$ grown through PLD could be employed in fabricating cost-effective optoelectronic devices, such as sensors and actuators whose responsiveness can be controlled by adjusting the laser spot along the sample's length and by varying the magnetic field's rotation across different planes of the sample.

\bibliography{reference}

\begin{thebibliography}{39}%
\makeatletter
\providecommand \@ifxundefined [1]{%
 \@ifx{#1\undefined}
}%
\providecommand \@ifnum [1]{%
 \ifnum #1\expandafter \@firstoftwo
 \else \expandafter \@secondoftwo
 \fi
}%
\providecommand \@ifx [1]{%
 \ifx #1\expandafter \@firstoftwo
 \else \expandafter \@secondoftwo
 \fi
}%
\providecommand \natexlab [1]{#1}%
\providecommand \enquote  [1]{``#1''}%
\providecommand \bibnamefont  [1]{#1}%
\providecommand \bibfnamefont [1]{#1}%
\providecommand \citenamefont [1]{#1}%
\providecommand \href@noop [0]{\@secondoftwo}%
\providecommand \href [0]{\begingroup \@sanitize@url \@href}%
\providecommand \@href[1]{\@@startlink{#1}\@@href}%
\providecommand \@@href[1]{\endgroup#1\@@endlink}%
\providecommand \@sanitize@url [0]{\catcode `\\12\catcode `\$12\catcode `\&12\catcode `\#12\catcode `\^12\catcode `\_12\catcode `\%12\relax}%
\providecommand \@@startlink[1]{}%
\providecommand \@@endlink[0]{}%
\providecommand \url  [0]{\begingroup\@sanitize@url \@url }%
\providecommand \@url [1]{\endgroup\@href {#1}{\urlprefix }}%
\providecommand \urlprefix  [0]{URL }%
\providecommand \Eprint [0]{\href }%
\providecommand \doibase [0]{https://doi.org/}%
\providecommand \selectlanguage [0]{\@gobble}%
\providecommand \bibinfo  [0]{\@secondoftwo}%
\providecommand \bibfield  [0]{\@secondoftwo}%
\providecommand \translation [1]{[#1]}%
\providecommand \BibitemOpen [0]{}%
\providecommand \bibitemStop [0]{}%
\providecommand \bibitemNoStop [0]{.\EOS\space}%
\providecommand \EOS [0]{\spacefactor3000\relax}%
\providecommand \BibitemShut  [1]{\csname bibitem#1\endcsname}%
\let\auto@bib@innerbib\@empty
\bibitem [{\citenamefont {Moore}(2010)}]{moore2010birth}%
  \BibitemOpen
  \bibfield  {author} {\bibinfo {author} {\bibfnamefont {J.~E.}\ \bibnamefont {Moore}},\ }\bibfield  {title} {\bibinfo {title} {The birth of topological insulators},\ }\href@noop {} {\bibfield  {journal} {\bibinfo  {journal} {Nature}\ }\textbf {\bibinfo {volume} {464}},\ \bibinfo {pages} {194} (\bibinfo {year} {2010})}\BibitemShut {NoStop}%
\bibitem [{\citenamefont {Zhang}\ \emph {et~al.}(2009)\citenamefont {Zhang}, \citenamefont {Liu}, \citenamefont {Qi}, \citenamefont {Dai}, \citenamefont {Fang},\ and\ \citenamefont {Zhang}}]{zhang2009topological}%
  \BibitemOpen
  \bibfield  {author} {\bibinfo {author} {\bibfnamefont {H.}~\bibnamefont {Zhang}}, \bibinfo {author} {\bibfnamefont {C.-X.}\ \bibnamefont {Liu}}, \bibinfo {author} {\bibfnamefont {X.-L.}\ \bibnamefont {Qi}}, \bibinfo {author} {\bibfnamefont {X.}~\bibnamefont {Dai}}, \bibinfo {author} {\bibfnamefont {Z.}~\bibnamefont {Fang}},\ and\ \bibinfo {author} {\bibfnamefont {S.-C.}\ \bibnamefont {Zhang}},\ }\bibfield  {title} {\bibinfo {title} {Topological insulators in bi2se3, bi2te3 and sb2te3 with a single dirac cone on the surface},\ }\href@noop {} {\bibfield  {journal} {\bibinfo  {journal} {Nature physics}\ }\textbf {\bibinfo {volume} {5}},\ \bibinfo {pages} {438} (\bibinfo {year} {2009})}\BibitemShut {NoStop}%
\bibitem [{\citenamefont {Hasan}\ and\ \citenamefont {Kane}(2010)}]{hasan2010colloquium}%
  \BibitemOpen
  \bibfield  {author} {\bibinfo {author} {\bibfnamefont {M.~Z.}\ \bibnamefont {Hasan}}\ and\ \bibinfo {author} {\bibfnamefont {C.~L.}\ \bibnamefont {Kane}},\ }\bibfield  {title} {\bibinfo {title} {Colloquium: topological insulators},\ }\href@noop {} {\bibfield  {journal} {\bibinfo  {journal} {Reviews of modern physics}\ }\textbf {\bibinfo {volume} {82}},\ \bibinfo {pages} {3045} (\bibinfo {year} {2010})}\BibitemShut {NoStop}%
\bibitem [{\citenamefont {Ando}(2013)}]{ando2013topological}%
  \BibitemOpen
  \bibfield  {author} {\bibinfo {author} {\bibfnamefont {Y.}~\bibnamefont {Ando}},\ }\bibfield  {title} {\bibinfo {title} {Topological insulator materials},\ }\href@noop {} {\bibfield  {journal} {\bibinfo  {journal} {Journal of the Physical Society of Japan}\ }\textbf {\bibinfo {volume} {82}},\ \bibinfo {pages} {102001} (\bibinfo {year} {2013})}\BibitemShut {NoStop}%
\bibitem [{\citenamefont {Hasan}\ and\ \citenamefont {Moore}(2011)}]{hasan2011three}%
  \BibitemOpen
  \bibfield  {author} {\bibinfo {author} {\bibfnamefont {M.~Z.}\ \bibnamefont {Hasan}}\ and\ \bibinfo {author} {\bibfnamefont {J.~E.}\ \bibnamefont {Moore}},\ }\bibfield  {title} {\bibinfo {title} {Three-dimensional topological insulators},\ }\href@noop {} {\bibfield  {journal} {\bibinfo  {journal} {Annu. Rev. Condens. Matter Phys.}\ }\textbf {\bibinfo {volume} {2}},\ \bibinfo {pages} {55} (\bibinfo {year} {2011})}\BibitemShut {NoStop}%
\bibitem [{\citenamefont {He}\ \emph {et~al.}(2019)\citenamefont {He}, \citenamefont {Sun},\ and\ \citenamefont {He}}]{he2019topological}%
  \BibitemOpen
  \bibfield  {author} {\bibinfo {author} {\bibfnamefont {M.}~\bibnamefont {He}}, \bibinfo {author} {\bibfnamefont {H.}~\bibnamefont {Sun}},\ and\ \bibinfo {author} {\bibfnamefont {Q.~L.}\ \bibnamefont {He}},\ }\bibfield  {title} {\bibinfo {title} {Topological insulator: Spintronics and quantum computations},\ }\href@noop {} {\bibfield  {journal} {\bibinfo  {journal} {Frontiers of Physics}\ }\textbf {\bibinfo {volume} {14}},\ \bibinfo {pages} {1} (\bibinfo {year} {2019})}\BibitemShut {NoStop}%
\bibitem [{\citenamefont {Politano}\ \emph {et~al.}(2017)\citenamefont {Politano}, \citenamefont {Viti},\ and\ \citenamefont {Vitiello}}]{politano2017optoelectronic}%
  \BibitemOpen
  \bibfield  {author} {\bibinfo {author} {\bibfnamefont {A.}~\bibnamefont {Politano}}, \bibinfo {author} {\bibfnamefont {L.}~\bibnamefont {Viti}},\ and\ \bibinfo {author} {\bibfnamefont {M.~S.}\ \bibnamefont {Vitiello}},\ }\bibfield  {title} {\bibinfo {title} {Optoelectronic devices, plasmonics, and photonics with topological insulators},\ }\href@noop {} {\bibfield  {journal} {\bibinfo  {journal} {APL Materials}\ }\textbf {\bibinfo {volume} {5}} (\bibinfo {year} {2017})}\BibitemShut {NoStop}%
\bibitem [{\citenamefont {McGuire}\ and\ \citenamefont {Potter}(1975)}]{1058782}%
  \BibitemOpen
  \bibfield  {author} {\bibinfo {author} {\bibfnamefont {T.}~\bibnamefont {McGuire}}\ and\ \bibinfo {author} {\bibfnamefont {R.}~\bibnamefont {Potter}},\ }\bibfield  {title} {\bibinfo {title} {Anisotropic magnetoresistance in ferromagnetic 3d alloys},\ }\href {https://doi.org/10.1109/TMAG.1975.1058782} {\bibfield  {journal} {\bibinfo  {journal} {IEEE Transactions on Magnetics}\ }\textbf {\bibinfo {volume} {11}},\ \bibinfo {pages} {1018} (\bibinfo {year} {1975})}\BibitemShut {NoStop}%
\bibitem [{\citenamefont {Ky}(1967)}]{ky1967planar}%
  \BibitemOpen
  \bibfield  {author} {\bibinfo {author} {\bibfnamefont {V.~D.}\ \bibnamefont {Ky}},\ }\bibfield  {title} {\bibinfo {title} {Planar hall and nernst effect in ferromagnetic metals},\ }\href@noop {} {\bibfield  {journal} {\bibinfo  {journal} {physica status solidi (b)}\ }\textbf {\bibinfo {volume} {22}},\ \bibinfo {pages} {729} (\bibinfo {year} {1967})}\BibitemShut {NoStop}%
\bibitem [{\citenamefont {Smit}(1951)}]{smit1951magnetoresistance}%
  \BibitemOpen
  \bibfield  {author} {\bibinfo {author} {\bibfnamefont {J.}~\bibnamefont {Smit}},\ }\bibfield  {title} {\bibinfo {title} {Magnetoresistance of ferromagnetic metals and alloys at low temperatures},\ }\href@noop {} {\bibfield  {journal} {\bibinfo  {journal} {Physica}\ }\textbf {\bibinfo {volume} {17}},\ \bibinfo {pages} {612} (\bibinfo {year} {1951})}\BibitemShut {NoStop}%
\bibitem [{\citenamefont {Campbell}\ \emph {et~al.}(1970)\citenamefont {Campbell}, \citenamefont {Fert},\ and\ \citenamefont {Jaoul}}]{campbell1970spontaneous}%
  \BibitemOpen
  \bibfield  {author} {\bibinfo {author} {\bibfnamefont {I.}~\bibnamefont {Campbell}}, \bibinfo {author} {\bibfnamefont {A.}~\bibnamefont {Fert}},\ and\ \bibinfo {author} {\bibfnamefont {O.}~\bibnamefont {Jaoul}},\ }\bibfield  {title} {\bibinfo {title} {The spontaneous resistivity anisotropy in ni-based alloys},\ }\href@noop {} {\bibfield  {journal} {\bibinfo  {journal} {Journal of Physics C: Solid State Physics}\ }\textbf {\bibinfo {volume} {3}},\ \bibinfo {pages} {S95} (\bibinfo {year} {1970})}\BibitemShut {NoStop}%
\bibitem [{\citenamefont {Taskin}\ \emph {et~al.}(2017)\citenamefont {Taskin}, \citenamefont {Legg}, \citenamefont {Yang}, \citenamefont {Sasaki}, \citenamefont {Kanai}, \citenamefont {Matsumoto}, \citenamefont {Rosch},\ and\ \citenamefont {Ando}}]{taskin2017planar}%
  \BibitemOpen
  \bibfield  {author} {\bibinfo {author} {\bibfnamefont {A.}~\bibnamefont {Taskin}}, \bibinfo {author} {\bibfnamefont {H.~F.}\ \bibnamefont {Legg}}, \bibinfo {author} {\bibfnamefont {F.}~\bibnamefont {Yang}}, \bibinfo {author} {\bibfnamefont {S.}~\bibnamefont {Sasaki}}, \bibinfo {author} {\bibfnamefont {Y.}~\bibnamefont {Kanai}}, \bibinfo {author} {\bibfnamefont {K.}~\bibnamefont {Matsumoto}}, \bibinfo {author} {\bibfnamefont {A.}~\bibnamefont {Rosch}},\ and\ \bibinfo {author} {\bibfnamefont {Y.}~\bibnamefont {Ando}},\ }\bibfield  {title} {\bibinfo {title} {Planar hall effect from the surface of topological insulators},\ }\href@noop {} {\bibfield  {journal} {\bibinfo  {journal} {Nature communications}\ }\textbf {\bibinfo {volume} {8}},\ \bibinfo {pages} {1340} (\bibinfo {year} {2017})}\BibitemShut {NoStop}%
\bibitem [{\citenamefont {Budhani}\ \emph {et~al.}(2021)\citenamefont {Budhani}, \citenamefont {Higgins}, \citenamefont {McAlmont},\ and\ \citenamefont {Paglione}}]{budhani2021planar}%
  \BibitemOpen
  \bibfield  {author} {\bibinfo {author} {\bibfnamefont {R.~C.}\ \bibnamefont {Budhani}}, \bibinfo {author} {\bibfnamefont {J.~S.}\ \bibnamefont {Higgins}}, \bibinfo {author} {\bibfnamefont {D.}~\bibnamefont {McAlmont}},\ and\ \bibinfo {author} {\bibfnamefont {J.}~\bibnamefont {Paglione}},\ }\bibfield  {title} {\bibinfo {title} {Planar hall effect in c-axis textured films of bi85sb15 topological insulator},\ }\href@noop {} {\bibfield  {journal} {\bibinfo  {journal} {AIP Advances}\ }\textbf {\bibinfo {volume} {11}} (\bibinfo {year} {2021})}\BibitemShut {NoStop}%
\bibitem [{\citenamefont {Kumar}\ \emph {et~al.}(2018)\citenamefont {Kumar}, \citenamefont {Guin}, \citenamefont {Felser},\ and\ \citenamefont {Shekhar}}]{kumar2018planar}%
  \BibitemOpen
  \bibfield  {author} {\bibinfo {author} {\bibfnamefont {N.}~\bibnamefont {Kumar}}, \bibinfo {author} {\bibfnamefont {S.~N.}\ \bibnamefont {Guin}}, \bibinfo {author} {\bibfnamefont {C.}~\bibnamefont {Felser}},\ and\ \bibinfo {author} {\bibfnamefont {C.}~\bibnamefont {Shekhar}},\ }\bibfield  {title} {\bibinfo {title} {Planar hall effect in the weyl semimetal gdptbi},\ }\href@noop {} {\bibfield  {journal} {\bibinfo  {journal} {Physical Review B}\ }\textbf {\bibinfo {volume} {98}},\ \bibinfo {pages} {041103} (\bibinfo {year} {2018})}\BibitemShut {NoStop}%
\bibitem [{\citenamefont {Li}\ \emph {et~al.}(2018)\citenamefont {Li}, \citenamefont {Wang}, \citenamefont {He}, \citenamefont {Wang},\ and\ \citenamefont {Shen}}]{li2018giant}%
  \BibitemOpen
  \bibfield  {author} {\bibinfo {author} {\bibfnamefont {H.}~\bibnamefont {Li}}, \bibinfo {author} {\bibfnamefont {H.-W.}\ \bibnamefont {Wang}}, \bibinfo {author} {\bibfnamefont {H.}~\bibnamefont {He}}, \bibinfo {author} {\bibfnamefont {J.}~\bibnamefont {Wang}},\ and\ \bibinfo {author} {\bibfnamefont {S.-Q.}\ \bibnamefont {Shen}},\ }\bibfield  {title} {\bibinfo {title} {Giant anisotropic magnetoresistance and planar hall effect in the dirac semimetal cd 3 as 2},\ }\href@noop {} {\bibfield  {journal} {\bibinfo  {journal} {Physical Review B}\ }\textbf {\bibinfo {volume} {97}},\ \bibinfo {pages} {201110} (\bibinfo {year} {2018})}\BibitemShut {NoStop}%
\bibitem [{\citenamefont {Nandy}\ \emph {et~al.}(2017)\citenamefont {Nandy}, \citenamefont {Sharma}, \citenamefont {Taraphder},\ and\ \citenamefont {Tewari}}]{nandy2017chiral}%
  \BibitemOpen
  \bibfield  {author} {\bibinfo {author} {\bibfnamefont {S.}~\bibnamefont {Nandy}}, \bibinfo {author} {\bibfnamefont {G.}~\bibnamefont {Sharma}}, \bibinfo {author} {\bibfnamefont {A.}~\bibnamefont {Taraphder}},\ and\ \bibinfo {author} {\bibfnamefont {S.}~\bibnamefont {Tewari}},\ }\bibfield  {title} {\bibinfo {title} {Chiral anomaly as the origin of the planar hall effect in weyl semimetals},\ }\href@noop {} {\bibfield  {journal} {\bibinfo  {journal} {Physical review letters}\ }\textbf {\bibinfo {volume} {119}},\ \bibinfo {pages} {176804} (\bibinfo {year} {2017})}\BibitemShut {NoStop}%
\bibitem [{\citenamefont {Sulaev}\ \emph {et~al.}(2015)\citenamefont {Sulaev}, \citenamefont {Zeng}, \citenamefont {Shen}, \citenamefont {Cho}, \citenamefont {Zhu}, \citenamefont {Feng}, \citenamefont {Eremeev}, \citenamefont {Kawazoe}, \citenamefont {Shen},\ and\ \citenamefont {Wang}}]{sulaev2015electrically}%
  \BibitemOpen
  \bibfield  {author} {\bibinfo {author} {\bibfnamefont {A.}~\bibnamefont {Sulaev}}, \bibinfo {author} {\bibfnamefont {M.}~\bibnamefont {Zeng}}, \bibinfo {author} {\bibfnamefont {S.-Q.}\ \bibnamefont {Shen}}, \bibinfo {author} {\bibfnamefont {S.~K.}\ \bibnamefont {Cho}}, \bibinfo {author} {\bibfnamefont {W.~G.}\ \bibnamefont {Zhu}}, \bibinfo {author} {\bibfnamefont {Y.~P.}\ \bibnamefont {Feng}}, \bibinfo {author} {\bibfnamefont {S.~V.}\ \bibnamefont {Eremeev}}, \bibinfo {author} {\bibfnamefont {Y.}~\bibnamefont {Kawazoe}}, \bibinfo {author} {\bibfnamefont {L.}~\bibnamefont {Shen}},\ and\ \bibinfo {author} {\bibfnamefont {L.}~\bibnamefont {Wang}},\ }\bibfield  {title} {\bibinfo {title} {Electrically tunable in-plane anisotropic magnetoresistance in topological insulator bisbtese2 nanodevices},\ }\href@noop {} {\bibfield  {journal} {\bibinfo  {journal} {Nano letters}\ }\textbf {\bibinfo {volume} {15}},\ \bibinfo {pages} {2061} (\bibinfo {year} {2015})}\BibitemShut {NoStop}%
\bibitem [{\citenamefont {Nandy}\ \emph {et~al.}(2018)\citenamefont {Nandy}, \citenamefont {Taraphder},\ and\ \citenamefont {Tewari}}]{nandy2018berry}%
  \BibitemOpen
  \bibfield  {author} {\bibinfo {author} {\bibfnamefont {S.}~\bibnamefont {Nandy}}, \bibinfo {author} {\bibfnamefont {A.}~\bibnamefont {Taraphder}},\ and\ \bibinfo {author} {\bibfnamefont {S.}~\bibnamefont {Tewari}},\ }\bibfield  {title} {\bibinfo {title} {Berry phase theory of planar hall effect in topological insulators},\ }\href@noop {} {\bibfield  {journal} {\bibinfo  {journal} {Scientific Reports}\ }\textbf {\bibinfo {volume} {8}},\ \bibinfo {pages} {14983} (\bibinfo {year} {2018})}\BibitemShut {NoStop}%
\bibitem [{\citenamefont {Akzyanov}\ and\ \citenamefont {Rakhmanov}(2018)}]{akzyanov2018surface}%
  \BibitemOpen
  \bibfield  {author} {\bibinfo {author} {\bibfnamefont {R.}~\bibnamefont {Akzyanov}}\ and\ \bibinfo {author} {\bibfnamefont {A.}~\bibnamefont {Rakhmanov}},\ }\bibfield  {title} {\bibinfo {title} {Surface charge conductivity of a topological insulator in a magnetic field: The effect of hexagonal warping},\ }\href@noop {} {\bibfield  {journal} {\bibinfo  {journal} {Physical Review B}\ }\textbf {\bibinfo {volume} {97}},\ \bibinfo {pages} {075421} (\bibinfo {year} {2018})}\BibitemShut {NoStop}%
\bibitem [{\citenamefont {Zheng}\ \emph {et~al.}(2020)\citenamefont {Zheng}, \citenamefont {Duan}, \citenamefont {Wang}, \citenamefont {Li}, \citenamefont {Deng},\ and\ \citenamefont {Wang}}]{zheng2020origin}%
  \BibitemOpen
  \bibfield  {author} {\bibinfo {author} {\bibfnamefont {S.-H.}\ \bibnamefont {Zheng}}, \bibinfo {author} {\bibfnamefont {H.-J.}\ \bibnamefont {Duan}}, \bibinfo {author} {\bibfnamefont {J.-K.}\ \bibnamefont {Wang}}, \bibinfo {author} {\bibfnamefont {J.-Y.}\ \bibnamefont {Li}}, \bibinfo {author} {\bibfnamefont {M.-X.}\ \bibnamefont {Deng}},\ and\ \bibinfo {author} {\bibfnamefont {R.-Q.}\ \bibnamefont {Wang}},\ }\bibfield  {title} {\bibinfo {title} {Origin of planar hall effect on the surface of topological insulators: Tilt of dirac cone by an in-plane magnetic field},\ }\href@noop {} {\bibfield  {journal} {\bibinfo  {journal} {Physical Review B}\ }\textbf {\bibinfo {volume} {101}},\ \bibinfo {pages} {041408} (\bibinfo {year} {2020})}\BibitemShut {NoStop}%
\bibitem [{\citenamefont {Imamuddin}\ and\ \citenamefont {Dupre}(1972)}]{imamuddin1972thermoelectric}%
  \BibitemOpen
  \bibfield  {author} {\bibinfo {author} {\bibfnamefont {M.}~\bibnamefont {Imamuddin}}\ and\ \bibinfo {author} {\bibfnamefont {A.}~\bibnamefont {Dupre}},\ }\bibfield  {title} {\bibinfo {title} {Thermoelectric properties of p-type bi2te3--sb2te3--sb2se3 alloys and n-type bi2te3--bi2se3 alloys in the temperature range 300 to 600 k},\ }\href@noop {} {\bibfield  {journal} {\bibinfo  {journal} {physica status solidi (a)}\ }\textbf {\bibinfo {volume} {10}},\ \bibinfo {pages} {415} (\bibinfo {year} {1972})}\BibitemShut {NoStop}%
\bibitem [{\citenamefont {Xu}\ \emph {et~al.}(2017)\citenamefont {Xu}, \citenamefont {Xu},\ and\ \citenamefont {Zhu}}]{xu2017topological}%
  \BibitemOpen
  \bibfield  {author} {\bibinfo {author} {\bibfnamefont {N.}~\bibnamefont {Xu}}, \bibinfo {author} {\bibfnamefont {Y.}~\bibnamefont {Xu}},\ and\ \bibinfo {author} {\bibfnamefont {J.}~\bibnamefont {Zhu}},\ }\bibfield  {title} {\bibinfo {title} {Topological insulators for thermoelectrics},\ }\href@noop {} {\bibfield  {journal} {\bibinfo  {journal} {npj Quantum Materials}\ }\textbf {\bibinfo {volume} {2}},\ \bibinfo {pages} {51} (\bibinfo {year} {2017})}\BibitemShut {NoStop}%
\bibitem [{\citenamefont {Ivanov}\ \emph {et~al.}(2018)\citenamefont {Ivanov}, \citenamefont {Burkov},\ and\ \citenamefont {Pshenay-Severin}}]{ivanov2018thermoelectric}%
  \BibitemOpen
  \bibfield  {author} {\bibinfo {author} {\bibfnamefont {Y.~V.}\ \bibnamefont {Ivanov}}, \bibinfo {author} {\bibfnamefont {A.~T.}\ \bibnamefont {Burkov}},\ and\ \bibinfo {author} {\bibfnamefont {D.~A.}\ \bibnamefont {Pshenay-Severin}},\ }\bibfield  {title} {\bibinfo {title} {Thermoelectric properties of topological insulators},\ }\href@noop {} {\bibfield  {journal} {\bibinfo  {journal} {physica status solidi (b)}\ }\textbf {\bibinfo {volume} {255}},\ \bibinfo {pages} {1800020} (\bibinfo {year} {2018})}\BibitemShut {NoStop}%
\bibitem [{\citenamefont {Osterhage}\ \emph {et~al.}(2014)\citenamefont {Osterhage}, \citenamefont {Gooth}, \citenamefont {Hamdou}, \citenamefont {Gwozdz}, \citenamefont {Zierold},\ and\ \citenamefont {Nielsch}}]{osterhage2014thermoelectric}%
  \BibitemOpen
  \bibfield  {author} {\bibinfo {author} {\bibfnamefont {H.}~\bibnamefont {Osterhage}}, \bibinfo {author} {\bibfnamefont {J.}~\bibnamefont {Gooth}}, \bibinfo {author} {\bibfnamefont {B.}~\bibnamefont {Hamdou}}, \bibinfo {author} {\bibfnamefont {P.}~\bibnamefont {Gwozdz}}, \bibinfo {author} {\bibfnamefont {R.}~\bibnamefont {Zierold}},\ and\ \bibinfo {author} {\bibfnamefont {K.}~\bibnamefont {Nielsch}},\ }\bibfield  {title} {\bibinfo {title} {Thermoelectric properties of topological insulator bi2te3, sb2te3, and bi2se3 thin film quantum wells},\ }\href@noop {} {\bibfield  {journal} {\bibinfo  {journal} {Applied Physics Letters}\ }\textbf {\bibinfo {volume} {105}} (\bibinfo {year} {2014})}\BibitemShut {NoStop}%
\bibitem [{\citenamefont {Das}\ \emph {et~al.}(2015)\citenamefont {Das}, \citenamefont {Malik}, \citenamefont {Deb}, \citenamefont {Dhara}, \citenamefont {Bandyopadhyay},\ and\ \citenamefont {Banerjee}}]{das2015defect}%
  \BibitemOpen
  \bibfield  {author} {\bibinfo {author} {\bibfnamefont {D.}~\bibnamefont {Das}}, \bibinfo {author} {\bibfnamefont {K.}~\bibnamefont {Malik}}, \bibinfo {author} {\bibfnamefont {A.}~\bibnamefont {Deb}}, \bibinfo {author} {\bibfnamefont {S.}~\bibnamefont {Dhara}}, \bibinfo {author} {\bibfnamefont {S.}~\bibnamefont {Bandyopadhyay}},\ and\ \bibinfo {author} {\bibfnamefont {A.}~\bibnamefont {Banerjee}},\ }\bibfield  {title} {\bibinfo {title} {Defect induced structural and thermoelectric properties of sb2te3 alloy},\ }\href@noop {} {\bibfield  {journal} {\bibinfo  {journal} {Journal of Applied Physics}\ }\textbf {\bibinfo {volume} {118}} (\bibinfo {year} {2015})}\BibitemShut {NoStop}%
\bibitem [{\citenamefont {Yu}\ \emph {et~al.}(2019)\citenamefont {Yu}, \citenamefont {Zhu},\ and\ \citenamefont {Su}}]{yu2019unidirectional}%
  \BibitemOpen
  \bibfield  {author} {\bibinfo {author} {\bibfnamefont {X.-Q.}\ \bibnamefont {Yu}}, \bibinfo {author} {\bibfnamefont {Z.-G.}\ \bibnamefont {Zhu}},\ and\ \bibinfo {author} {\bibfnamefont {G.}~\bibnamefont {Su}},\ }\bibfield  {title} {\bibinfo {title} {Unidirectional seebeck effect in magnetic topological insulators},\ }\href@noop {} {\bibfield  {journal} {\bibinfo  {journal} {Physical Review B}\ }\textbf {\bibinfo {volume} {100}},\ \bibinfo {pages} {195418} (\bibinfo {year} {2019})}\BibitemShut {NoStop}%
\bibitem [{\citenamefont {Yu}\ \emph {et~al.}(2021)\citenamefont {Yu}, \citenamefont {Zhu},\ and\ \citenamefont {Su}}]{yu2021hexagonal}%
  \BibitemOpen
  \bibfield  {author} {\bibinfo {author} {\bibfnamefont {X.-Q.}\ \bibnamefont {Yu}}, \bibinfo {author} {\bibfnamefont {Z.-G.}\ \bibnamefont {Zhu}},\ and\ \bibinfo {author} {\bibfnamefont {G.}~\bibnamefont {Su}},\ }\bibfield  {title} {\bibinfo {title} {Hexagonal warping induced nonlinear planar nernst effect in nonmagnetic topological insulators},\ }\href@noop {} {\bibfield  {journal} {\bibinfo  {journal} {Physical Review B}\ }\textbf {\bibinfo {volume} {103}},\ \bibinfo {pages} {035410} (\bibinfo {year} {2021})}\BibitemShut {NoStop}%
\bibitem [{\citenamefont {Jain}\ \emph {et~al.}(2022)\citenamefont {Jain}, \citenamefont {Stanley}, \citenamefont {Bose}, \citenamefont {Richardella}, \citenamefont {Zhang}, \citenamefont {Pillsbury}, \citenamefont {Muller}, \citenamefont {Samarth},\ and\ \citenamefont {Ralph}}]{jain2022thermally}%
  \BibitemOpen
  \bibfield  {author} {\bibinfo {author} {\bibfnamefont {R.}~\bibnamefont {Jain}}, \bibinfo {author} {\bibfnamefont {M.}~\bibnamefont {Stanley}}, \bibinfo {author} {\bibfnamefont {A.}~\bibnamefont {Bose}}, \bibinfo {author} {\bibfnamefont {A.~R.}\ \bibnamefont {Richardella}}, \bibinfo {author} {\bibfnamefont {X.~S.}\ \bibnamefont {Zhang}}, \bibinfo {author} {\bibfnamefont {T.}~\bibnamefont {Pillsbury}}, \bibinfo {author} {\bibfnamefont {D.~A.}\ \bibnamefont {Muller}}, \bibinfo {author} {\bibfnamefont {N.}~\bibnamefont {Samarth}},\ and\ \bibinfo {author} {\bibfnamefont {D.~C.}\ \bibnamefont {Ralph}},\ }\bibfield  {title} {\bibinfo {title} {Thermally-generated spin current in the topological insulator bi $ \_2 $ se $ \_3$},\ }\href@noop {} {\bibfield  {journal} {\bibinfo  {journal} {arXiv preprint arXiv:2210.05636}\ } (\bibinfo {year} {2022})}\BibitemShut {NoStop}%
\bibitem [{\citenamefont {Aono}(1970)}]{aono1970nernst}%
  \BibitemOpen
  \bibfield  {author} {\bibinfo {author} {\bibfnamefont {T.}~\bibnamefont {Aono}},\ }\bibfield  {title} {\bibinfo {title} {The nernst effect of bi-sb alloys},\ }\href@noop {} {\bibfield  {journal} {\bibinfo  {journal} {Japanese Journal of Applied Physics}\ }\textbf {\bibinfo {volume} {9}},\ \bibinfo {pages} {761} (\bibinfo {year} {1970})}\BibitemShut {NoStop}%
\bibitem [{\citenamefont {Barnard}\ and\ \citenamefont {Cannella}(1974)}]{barnard1974thermoelectricity}%
  \BibitemOpen
  \bibfield  {author} {\bibinfo {author} {\bibfnamefont {R.~D.}\ \bibnamefont {Barnard}}\ and\ \bibinfo {author} {\bibfnamefont {V.}~\bibnamefont {Cannella}},\ }\bibfield  {title} {\bibinfo {title} {Thermoelectricity in metals and alloys},\ }\href@noop {} {\bibfield  {journal} {\bibinfo  {journal} {Physics Today}\ }\textbf {\bibinfo {volume} {27}},\ \bibinfo {pages} {52} (\bibinfo {year} {1974})}\BibitemShut {NoStop}%
\bibitem [{\citenamefont {McIver}\ \emph {et~al.}(2012)\citenamefont {McIver}, \citenamefont {Hsieh}, \citenamefont {Steinberg}, \citenamefont {Jarillo-Herrero},\ and\ \citenamefont {Gedik}}]{mciver2012control}%
  \BibitemOpen
  \bibfield  {author} {\bibinfo {author} {\bibfnamefont {J.}~\bibnamefont {McIver}}, \bibinfo {author} {\bibfnamefont {D.}~\bibnamefont {Hsieh}}, \bibinfo {author} {\bibfnamefont {H.}~\bibnamefont {Steinberg}}, \bibinfo {author} {\bibfnamefont {P.}~\bibnamefont {Jarillo-Herrero}},\ and\ \bibinfo {author} {\bibfnamefont {N.}~\bibnamefont {Gedik}},\ }\bibfield  {title} {\bibinfo {title} {Control over topological insulator photocurrents with light polarization},\ }\href@noop {} {\bibfield  {journal} {\bibinfo  {journal} {Nature nanotechnology}\ }\textbf {\bibinfo {volume} {7}},\ \bibinfo {pages} {96} (\bibinfo {year} {2012})}\BibitemShut {NoStop}%
\bibitem [{\citenamefont {Pan}\ \emph {et~al.}(2017)\citenamefont {Pan}, \citenamefont {Wang}, \citenamefont {Yeats}, \citenamefont {Pillsbury}, \citenamefont {Flanagan}, \citenamefont {Richardella}, \citenamefont {Zhang}, \citenamefont {Awschalom}, \citenamefont {Liu},\ and\ \citenamefont {Samarth}}]{pan2017helicity}%
  \BibitemOpen
  \bibfield  {author} {\bibinfo {author} {\bibfnamefont {Y.}~\bibnamefont {Pan}}, \bibinfo {author} {\bibfnamefont {Q.-Z.}\ \bibnamefont {Wang}}, \bibinfo {author} {\bibfnamefont {A.~L.}\ \bibnamefont {Yeats}}, \bibinfo {author} {\bibfnamefont {T.}~\bibnamefont {Pillsbury}}, \bibinfo {author} {\bibfnamefont {T.~C.}\ \bibnamefont {Flanagan}}, \bibinfo {author} {\bibfnamefont {A.}~\bibnamefont {Richardella}}, \bibinfo {author} {\bibfnamefont {H.}~\bibnamefont {Zhang}}, \bibinfo {author} {\bibfnamefont {D.~D.}\ \bibnamefont {Awschalom}}, \bibinfo {author} {\bibfnamefont {C.-X.}\ \bibnamefont {Liu}},\ and\ \bibinfo {author} {\bibfnamefont {N.}~\bibnamefont {Samarth}},\ }\bibfield  {title} {\bibinfo {title} {Helicity dependent photocurrent in electrically gated (bi1- x sb x) 2te3 thin films},\ }\href@noop {} {\bibfield  {journal} {\bibinfo  {journal} {Nature communications}\ }\textbf {\bibinfo {volume} {8}},\ \bibinfo {pages} {1037} (\bibinfo {year} {2017})}\BibitemShut {NoStop}%
\bibitem [{\citenamefont {Roy}\ \emph {et~al.}(2022)\citenamefont {Roy}, \citenamefont {Manna}, \citenamefont {Mitra},\ and\ \citenamefont {Pal}}]{roy2022photothermal}%
  \BibitemOpen
  \bibfield  {author} {\bibinfo {author} {\bibfnamefont {S.}~\bibnamefont {Roy}}, \bibinfo {author} {\bibfnamefont {S.}~\bibnamefont {Manna}}, \bibinfo {author} {\bibfnamefont {C.}~\bibnamefont {Mitra}},\ and\ \bibinfo {author} {\bibfnamefont {B.}~\bibnamefont {Pal}},\ }\bibfield  {title} {\bibinfo {title} {Photothermal control of helicity-dependent current in epitaxial sb2te2se topological insulator thin-films at ambient temperature},\ }\href@noop {} {\bibfield  {journal} {\bibinfo  {journal} {ACS Applied Materials \& Interfaces}\ }\textbf {\bibinfo {volume} {14}},\ \bibinfo {pages} {9909} (\bibinfo {year} {2022})}\BibitemShut {NoStop}%
\bibitem [{\citenamefont {Weber}\ \emph {et~al.}(2008)\citenamefont {Weber}, \citenamefont {Seidl}, \citenamefont {Bel’kov}, \citenamefont {Golub}, \citenamefont {Danilov}, \citenamefont {Ivchenko}, \citenamefont {Prettl}, \citenamefont {Kvon}, \citenamefont {Cho}, \citenamefont {Lee} \emph {et~al.}}]{weber2008magneto}%
  \BibitemOpen
  \bibfield  {author} {\bibinfo {author} {\bibfnamefont {W.}~\bibnamefont {Weber}}, \bibinfo {author} {\bibfnamefont {S.}~\bibnamefont {Seidl}}, \bibinfo {author} {\bibfnamefont {V.}~\bibnamefont {Bel’kov}}, \bibinfo {author} {\bibfnamefont {L.}~\bibnamefont {Golub}}, \bibinfo {author} {\bibfnamefont {S.}~\bibnamefont {Danilov}}, \bibinfo {author} {\bibfnamefont {E.}~\bibnamefont {Ivchenko}}, \bibinfo {author} {\bibfnamefont {W.}~\bibnamefont {Prettl}}, \bibinfo {author} {\bibfnamefont {Z.}~\bibnamefont {Kvon}}, \bibinfo {author} {\bibfnamefont {H.-I.}\ \bibnamefont {Cho}}, \bibinfo {author} {\bibfnamefont {J.-H.}\ \bibnamefont {Lee}}, \emph {et~al.},\ }\bibfield  {title} {\bibinfo {title} {Magneto-gyrotropic photogalvanic effects in gan/algan two-dimensional systems},\ }\href@noop {} {\bibfield  {journal} {\bibinfo  {journal} {Solid state communications}\ }\textbf {\bibinfo {volume} {145}},\ \bibinfo {pages} {56} (\bibinfo {year} {2008})}\BibitemShut {NoStop}%
\bibitem [{\citenamefont {Junck}\ \emph {et~al.}(2013)\citenamefont {Junck}, \citenamefont {Refael},\ and\ \citenamefont {von Oppen}}]{junck2013photocurrent}%
  \BibitemOpen
  \bibfield  {author} {\bibinfo {author} {\bibfnamefont {A.}~\bibnamefont {Junck}}, \bibinfo {author} {\bibfnamefont {G.}~\bibnamefont {Refael}},\ and\ \bibinfo {author} {\bibfnamefont {F.}~\bibnamefont {von Oppen}},\ }\bibfield  {title} {\bibinfo {title} {Photocurrent response of topological insulator surface states},\ }\href@noop {} {\bibfield  {journal} {\bibinfo  {journal} {Physical Review B}\ }\textbf {\bibinfo {volume} {88}},\ \bibinfo {pages} {075144} (\bibinfo {year} {2013})}\BibitemShut {NoStop}%
\bibitem [{\citenamefont {Chen}\ \emph {et~al.}(2021)\citenamefont {Chen}, \citenamefont {Yu}, \citenamefont {Zhu}, \citenamefont {Zeng}, \citenamefont {Chen}, \citenamefont {Liu}, \citenamefont {Zhang}, \citenamefont {Cheng},\ and\ \citenamefont {He}}]{chen2021plane}%
  \BibitemOpen
  \bibfield  {author} {\bibinfo {author} {\bibfnamefont {S.}~\bibnamefont {Chen}}, \bibinfo {author} {\bibfnamefont {J.}~\bibnamefont {Yu}}, \bibinfo {author} {\bibfnamefont {K.}~\bibnamefont {Zhu}}, \bibinfo {author} {\bibfnamefont {X.}~\bibnamefont {Zeng}}, \bibinfo {author} {\bibfnamefont {Y.}~\bibnamefont {Chen}}, \bibinfo {author} {\bibfnamefont {Y.}~\bibnamefont {Liu}}, \bibinfo {author} {\bibfnamefont {Y.}~\bibnamefont {Zhang}}, \bibinfo {author} {\bibfnamefont {S.}~\bibnamefont {Cheng}},\ and\ \bibinfo {author} {\bibfnamefont {K.}~\bibnamefont {He}},\ }\bibfield  {title} {\bibinfo {title} {In-plane magnetic field induced helicity dependent photogalvanic effect on the surface states of topological insulators (bixsb1- x) 2te3},\ }\href@noop {} {\bibfield  {journal} {\bibinfo  {journal} {Journal of Applied Physics}\ }\textbf {\bibinfo {volume} {130}} (\bibinfo {year} {2021})}\BibitemShut {NoStop}%
\bibitem [{\citenamefont {Morimoto}\ \emph {et~al.}(2016)\citenamefont {Morimoto}, \citenamefont {Zhong}, \citenamefont {Orenstein},\ and\ \citenamefont {Moore}}]{morimoto2016semiclassical}%
  \BibitemOpen
  \bibfield  {author} {\bibinfo {author} {\bibfnamefont {T.}~\bibnamefont {Morimoto}}, \bibinfo {author} {\bibfnamefont {S.}~\bibnamefont {Zhong}}, \bibinfo {author} {\bibfnamefont {J.}~\bibnamefont {Orenstein}},\ and\ \bibinfo {author} {\bibfnamefont {J.~E.}\ \bibnamefont {Moore}},\ }\bibfield  {title} {\bibinfo {title} {Semiclassical theory of nonlinear magneto-optical responses with applications to topological dirac/weyl semimetals},\ }\href@noop {} {\bibfield  {journal} {\bibinfo  {journal} {Physical Review B}\ }\textbf {\bibinfo {volume} {94}},\ \bibinfo {pages} {245121} (\bibinfo {year} {2016})}\BibitemShut {NoStop}%
\bibitem [{\citenamefont {San~Li}\ and\ \citenamefont {Rabson}(1970)}]{san1970nernst}%
  \BibitemOpen
  \bibfield  {author} {\bibinfo {author} {\bibfnamefont {S.}~\bibnamefont {San~Li}}\ and\ \bibinfo {author} {\bibfnamefont {T.}~\bibnamefont {Rabson}},\ }\bibfield  {title} {\bibinfo {title} {The nernst and the seebeck effects in te-doped bisb alloys},\ }\href@noop {} {\bibfield  {journal} {\bibinfo  {journal} {Solid-State Electronics}\ }\textbf {\bibinfo {volume} {13}},\ \bibinfo {pages} {153} (\bibinfo {year} {1970})}\BibitemShut {NoStop}%
\bibitem [{\citenamefont {Kim}\ \emph {et~al.}(2017)\citenamefont {Kim}, \citenamefont {Jeon}, \citenamefont {Choi}, \citenamefont {Lee}, \citenamefont {Surabhi}, \citenamefont {Jeong}, \citenamefont {Lee},\ and\ \citenamefont {Park}}]{kim2017observation}%
  \BibitemOpen
  \bibfield  {author} {\bibinfo {author} {\bibfnamefont {D.-J.}\ \bibnamefont {Kim}}, \bibinfo {author} {\bibfnamefont {C.-Y.}\ \bibnamefont {Jeon}}, \bibinfo {author} {\bibfnamefont {J.-G.}\ \bibnamefont {Choi}}, \bibinfo {author} {\bibfnamefont {J.~W.}\ \bibnamefont {Lee}}, \bibinfo {author} {\bibfnamefont {S.}~\bibnamefont {Surabhi}}, \bibinfo {author} {\bibfnamefont {J.-R.}\ \bibnamefont {Jeong}}, \bibinfo {author} {\bibfnamefont {K.-J.}\ \bibnamefont {Lee}},\ and\ \bibinfo {author} {\bibfnamefont {B.-G.}\ \bibnamefont {Park}},\ }\bibfield  {title} {\bibinfo {title} {Observation of transverse spin nernst magnetoresistance induced by thermal spin current in ferromagnet/non-magnet bilayers},\ }\href@noop {} {\bibfield  {journal} {\bibinfo  {journal} {Nature communications}\ }\textbf {\bibinfo {volume} {8}},\ \bibinfo {pages} {1400} (\bibinfo {year} {2017})}\BibitemShut {NoStop}%
\end{thebibliography}%
  \end{document}